\documentclass{iau_FM}
\usepackage{graphicx}
\title[FM 16.~~Cool stars in the HRD]
{Cool Stars\\ in the Hertzsprung--Russell Diagram}
\author[Jacco Th. van Loon]
{Jacco Th. van Loon$^1$}
\affiliation{$^1$Lennard-Jones Laboratories, Keele University, ST5 5BG, UK\\
                 email: {\tt j.t.van.loon@keele.ac.uk}}
\pubyear{2015}
\jname{Astronomy in Focus, Volume 1} 
\editors{Piero Benvenuti, ed.}
\begin{document}
\maketitle
\begin{abstract}
As the opening review to the focus meeting ``Stellar Behemoths: Red
Supergiants across the Local Universe'', I here provide a brief introduction
to red supergiants, setting the stage for subsequent contributions. I
highlight some recent activity in the field, and identify areas of progress,
areas where progress is needed, and how such progress might be achieved.
\end{abstract}
\firstsection
\section{Introduction}
Rooted in the book of Job, ``Behemoth'' is a worthy accolade to bestow upon
the class of red supergiants. This finds reason in the popular use of the
word. Thinking of the namesake Polish death metal band, red supergiants are
dying stars, and their explosive death sparks the production and distribution
of metals. In the cult game Final Fantasy, behemoths are beasts most feared
for their magic spell ``meteor''; indeed, red supergiants fabricate dust
grains, a process enshrouded in mystery. Princeton University's Farlex WordNet
describes behemoths as ``abnormally large and powerful'', ``unusual'', ``of
exceptional importance and reputation'', and that ``strongly influence the
course of events'' -- all applicable to red supergiants. And they are, simply
put, ``whoppers''.

Descendent from stars with birth masses in the region of 8--30 M$_\odot$, red
supergiants become large, hundreds of R$_\odot$ up to more than a thousand
R$_\odot$, because their mantles become convective. The reason they become
convective is because the energy production rate in the core can no longer be
balanced by radiative diffusion and instead bulk motion is induced to
transport the energy to the surface. This causes a drastic restructuring of
the mantle, with a much shallower density gradient. It becomes optically thin
at much larger radial distance from the core, and hence the temperature of
the plasma where this happens is relatively low, $T_{\rm eff}\sim3500$--4000 K,
such that equilibrium is maintained ($L_{\rm emitted}=L_{\rm produced}$).

To put this in perspective, a massive main sequence star would easily fit
within the orbit of Mercury, but its red supergiant progeny could swallow
Jupiter. At that point, it would take more than a decade to travel around its
surface at a speed of 10 km s$^{-1}$, which is little more than the thermal
motions in the star's atmosphere but only a few times smaller than the escape
speed from its surface. Hence red supergiants oscillate on timescales of
years, and their winds have speeds of just a few tens of km s$^{-1}$ taking
tens of thousands of years for matter to move out to one parsec distance. This
means that we may be able to read the immediate past of the red supergiant --
even if it is no longer there -- in the story told by its circumstellar
medium. The low escape speed makes it easy for these stars to lose mass, but
it critically depends on how long they spend their time doing this, how much
mass they eventually will have lost. It also means that rotation at the mere
km s$^{-1}$ level may already affect these stars.

Red supergiants matter, as actors and probes in processes that drive galaxy
evolution. But there are some important aspects of their formation, evolution
and behaviour that need to be better understood:\\

\begin{center}
{\footnotesize
\begin{tabular}{lcl}
\hline
importance of red supergiants & & related issues to be resolved \\
\hline
they return gas (and form dust) &
$\rightarrow$ & how much? \\
they lock up baryonic matter in remnants &
$\rightarrow$ & how much? \\
they are supernova progenitors &
$\rightarrow$ & which ones cause what type? \\
they are signposts of young populations &
$\rightarrow$ & what are their birth masses and lifetimes? \\
they stand out at infrared wavelengths &
$\rightarrow$ & how cool and bright do they really become? \\
they exhibit physics we do not understand &
$\rightarrow$ & convection, mass loss, binarity, $...$ \\
\hline
\end{tabular}
}
\end{center}

\section{Recent activity and progress in the field}

Stellar evolutionary models are becoming increasingly sophisticated. Yet none
of them include all physics, e.g.\ either rotation {\it or} magnetic fields,
and they often adopt different prescriptions for mixing (in particular the
degree of ``overshoot''). Even seemingly similar models can give rather
different results, especially below 10\,000 K -- the realm of the yellow
hypergiants and red supergiants (\cite[Martins \& Palacios
2013]{MartinsPalacios2013}). There is no agreement on the boundaries or
overlap with the lower-mass stars on the asymptotic giant branch (AGB),
exacerbated by the uncertainty regarding the super-AGB stars which are in
many ways akin to AGB stars but which do continue to burn carbon in their
core, and of which the ultimate fate -- electron-capture supernova or white
dwarf -- remains elusive.

One way in which the model results differ is in the occurrence and extent of
the blue loops, red supergiants becoming blue supergiant again before either
exploding or becoming red supergiant for a second time. Of course, SN\,1987A
is exactly such case, having surprised everyone by being blue at the time of
explosion yet displaying a ring of material as a wink to its red supergiant
past. Dust was produced in it (\cite[Matsuura \etal\ 2011]{MatsuuraEtal2011};
\cite[Indebetouw \etal\ 2014]{IndebetouwEtal2014}) but, ultimately, more dust
-- be it produced in the supernova ejecta, red supergiant or present in the
nearby interstellar medium (ISM) -- will have been destroyed by the supernova
remnant (\cite[Laki\'cevi\'c \etal\ 2015]{LakicevicEtal2015}; \cite[Temim
\etal\ 2015]{TemimEtal2015}). Theory indicates that rotation can encourage the
migration of red supergiants to the blue part in the Hertzsprung--Russell
diagram (\cite[Saio, Georgy \& Meynet 2013]{SaioGeorgyMeynet2013}).

SN\,1987A may once have looked much like the nearby red supergiant Betelgeuse
(\cite[van Loon 2013]{Vanloon2013}). It is so close that we can see the
convection cells bubbling up at its surface (\cite[Haubois \etal\
2009]{HauboisEtal2009}), the clouds form in its extended atmosphere
(\cite[Kervella \etal\ 2009]{KervellaEtal2009},
\cite[2011]{KervellaEtal2011}), and the effects of it -- and of its progenitor
-- on the local ISM (\cite[Decin \etal\ 2012]{DecinEtal2012}).

Red supergiants lose mass at rates of $10^{-6}$ to $10^{-4}$ M$_\odot$ yr$^{-1}$,
similar to -- or in excess of -- the nuclear burning rates (\cite[van Loon
\etal\ 1999]{VanloonEtal1999}, \cite[2005]{VanloonEtal2005}). This means that
the exact rate of mass loss will determine the lifetime of the red supergiant
phase as well as the total mass that will have been lost (\cite[Meynet \etal\
2015]{MeynetEtal2015}). Mass loss may happen mostly in intense, brief phases
either at the very end of red supergiant evolution, or before or after in the
eruptions of (not so) luminous blue variables (\cite[Rest \etal\
2012]{RestEtal2012}). Clearly, previous mass loss will affect the red
supergiant phase, and mass loss during the red supergiant phase will affect
subsequent evolution and the characteristics of the ensuing supernova as it
proceeds through the circumburst environment (\cite[van Loon
2010]{Vanloon2010}).

One of the most important breakthroughs in stellar evolution observations has
been the identification of the progenitors of supernov{\ae}. This has been
limited almost exclusively to the common type II plateau variety, confirming
that they arise from red supergiants with birth masses in the range 8--17
M$_\odot$ (\cite[Smartt 2015]{Smartt2015}). What happens to more massive red
supergiants remains a mystery.

Studies of extra-galactic populations of red supergiants have enabled progress
in our understanding of stellar evolution (\cite[Levesque \etal\
2006]{LevesqueEtal2006}, \cite[2007]{LevesqueEtal2007}) and stellar mass loss,
and in reconstructing the star formation history and mapping the stellar
feedback in spiral galaxies (\cite[Javadi \etal\ 2015]{JavadiEtal2015}). This
will benefit from facilities such as JWST and E-ELT.

\section{Concluding remarks -- how to move forward?}
Given the above, several areas of high priority emerge, defining possible
strategies for progress. These are not limited to the following short list
that I would like to advocate:\\

\noindent
\begin{tabular}{lcl}
1 & -- & theoretical studies of mass loss mechanisms; \\
2 & -- & surveys of the red-to-blue supergiant ratio, including pulsation
         characteristics and \\
  &    & surface abundances; \\
3 & -- & (continued) surveys for supernova progenitors; \\
4 & -- & determine the red supergiant luminosity functions in galaxies with
         differing metal \\
  &    & content and recent (spatially varying) star formation histories; \\
5 & -- & high resolution, high precision case studies in combination with
         three-dimensional\\
  &    & computational models. \\
\end{tabular}\\

\noindent
And finally, if close binary interaction does affect the evolution of a large
fraction of the massive star population (\cite[de Mink \etal\
(2014)]{DeminkEtal2014}, then observational and theoretical efforts must be
promoted that aim to quantify these effects.

\end{document}